\documentstyle[12pt,epsfig,a4,cite,epic,ecltree,subfigure,fancybox,floatfig]{article}
\begin{document}
\initfloatingfigs
\normalsize
\newpage
\pagestyle{plain}
\setcounter{page}{1}
\pagenumbering{arabic}
\normalsize
\setlength{\topmargin}{-1cm}
\setlength{\oddsidemargin}{-0.5cm}
\def\gama1{\gamma_{1}}
\begin{titlepage}
\normalsize
\begin{center}

{ EUROPEAN ORGANIZATION FOR NUCLEAR RESEARCH (CERN) }\\
\end{center}
\bigskip
\bigskip
\bigskip
\begin{center}
\bigskip
\bigskip
\bigskip \bigskip
{\Large \bf \boldmath {Search for $\gamma\gamma$ decays of a Higgs boson
produced in association with a fermion pair in
$\mathrm{e^+e^-}$ collisions at LEP}\unboldmath \\}
 
\bigskip \bigskip
\bigskip \bigskip
{\bf The  ALEPH Collaboration\footnote{)See next pages for the list of authors}) \\}
\bigskip
\bigskip \bigskip
\bigskip \bigskip
\bigskip \bigskip
\bigskip \bigskip
{\bf Abstract}
\bigskip
\end{center}
 
A search for $\gamma\gamma$ decays of a Higgs boson is 
performed in the data sample collected at LEP with the ALEPH detector 
between 1991 and 
1999. This corresponds to an integrated luminosity of
672~\,$\mathrm{pb}^{-1}$ at 
centre-of-mass energies ranging from 88 to 202~\,GeV.
The search is based on topologies arising from a Higgs boson produced
in association with a fermion 
pair via the Higgs-strahlung process $\mathrm{e^+e^-\to Hf\bar{f}}$, with
$\mathrm{f\bar{f}=\nu\bar\nu,\ e^+e^-,\mu^+\mu^-,\tau^+\tau^-}$ or $\mathrm{q\bar{q}}$.
Twenty-two events are selected in the data, while 28 events are
expected from standard model processes. An upper limit is derived,
as a function of the Higgs boson mass, 
on the product of the  $\mathrm{e^+e^-\to Hf\bar{f}}$ cross section
and the $\mathrm{H \to \gamma \gamma}$ branching fraction.
In particular, a fermiophobic Higgs boson produced
with the standard model cross section is excluded at $95\%$  
confidence level for all masses below $100.7\,\mathrm{GeV/{\it c}^2}$.

\bigskip\bigskip\bigskip
{ }
\vfill
\end{titlepage}

\pagestyle{empty}
\newpage
\small
%
%
\newlength{\saveparskip}
\newlength{\savetextheight}
\newlength{\savetopmargin}
\newlength{\savetextwidth}
\newlength{\saveoddsidemargin}
\newlength{\savetopsep}
\setlength{\saveparskip}{\parskip}
\setlength{\savetextheight}{\textheight}
\setlength{\savetopmargin}{\topmargin}
\setlength{\savetextwidth}{\textwidth}
\setlength{\saveoddsidemargin}{\oddsidemargin}
\setlength{\savetopsep}{\topsep}
%
%
\setlength{\parskip}{0.0cm}
\setlength{\textheight}{25.0cm}
\setlength{\topmargin}{-1.5cm}
\setlength{\textwidth}{16 cm}
\setlength{\oddsidemargin}{-0.0cm}
\setlength{\topsep}{1mm}
\pretolerance=10000
\centerline{\large\bf The ALEPH Collaboration}
\footnotesize
\vspace{0.5cm}
{\raggedbottom
\begin{sloppypar}
\samepage\noindent
R.~Barate,
D.~Decamp,
P.~Ghez,
C.~Goy,
S.~Jezequel,
J.-P.~Lees,
F.~Martin,
E.~Merle,
\mbox{M.-N.~Minard},
B.~Pietrzyk
\nopagebreak
\begin{center}
\parbox{15.5cm}{\sl\samepage
Laboratoire de Physique des Particules (LAPP), IN$^{2}$P$^{3}$-CNRS,
F-74019 Annecy-le-Vieux Cedex, France}
\end{center}\end{sloppypar}
\vspace{2mm}
\begin{sloppypar}
\noindent
S.~Bravo,
M.P.~Casado,
M.~Chmeissani,
J.M.~Crespo,
E.~Fernandez,
M.~Fernandez-Bosman,
Ll.~Garrido,$^{15}$
E.~Graug\'{e}s,
J.~Lopez,
M.~Martinez,
G.~Merino,
R.~Miquel,
Ll.M.~Mir,
A.~Pacheco,
D.~Paneque,
H.~Ruiz
\nopagebreak
\begin{center}
\parbox{15.5cm}{\sl\samepage
Institut de F\'{i}sica d'Altes Energies, Universitat Aut\`{o}noma
de Barcelona, E-08193 Bellaterra (Barcelona), Spain$^{7}$}
\end{center}\end{sloppypar}
\vspace{2mm}
\begin{sloppypar}
\noindent
A.~Colaleo,
D.~Creanza,
N.~De~Filippis,
M.~de~Palma,
G.~Iaselli,
G.~Maggi,
M.~Maggi,
S.~Nuzzo,
A.~Ranieri,
G.~Raso,
F.~Ruggieri,
G.~Selvaggi,
L.~Silvestris,
P.~Tempesta,
A.~Tricomi,$^{3}$
G.~Zito
\nopagebreak
\begin{center}
\parbox{15.5cm}{\sl\samepage
Dipartimento di Fisica, INFN Sezione di Bari, I-70126 Bari, Italy}
\end{center}\end{sloppypar}
\vspace{2mm}
\begin{sloppypar}
\noindent
X.~Huang,
J.~Lin,
Q. Ouyang,
T.~Wang,
Y.~Xie,
R.~Xu,
S.~Xue,
J.~Zhang,
L.~Zhang,
W.~Zhao
\nopagebreak
\begin{center}
\parbox{15.5cm}{\sl\samepage
Institute of High Energy Physics, Academia Sinica, Beijing, The People's
Republic of China$^{8}$}
\end{center}\end{sloppypar}
\vspace{2mm}
\begin{sloppypar}
\noindent
D.~Abbaneo,
G.~Boix,$^{6}$
O.~Buchm\"uller,
M.~Cattaneo,
F.~Cerutti,
G.~Dissertori,
H.~Drevermann,
R.W.~Forty,
M.~Frank,
F.~Gianotti,
T.C.~Greening,
A.W.~Halley,
J.B.~Hansen,
J.~Harvey,
P.~Janot,
B.~Jost,
M.~Kado,
V.~Lemaitre,
P.~Maley,
P.~Mato,
A.~Minten,
A.~Moutoussi,
F.~Ranjard,
L.~Rolandi,
D.~Schlatter,
M.~Schmitt,$^{20}$
O.~Schneider,$^{2}$
P.~Spagnolo,
W.~Tejessy,
F.~Teubert,
E.~Tournefier,
A.~Valassi,
J.J.~Ward,
A.E.~Wright
\nopagebreak
\begin{center}
\parbox{15.5cm}{\sl\samepage
European Laboratory for Particle Physics (CERN), CH-1211 Geneva 23,
Switzerland}
\end{center}\end{sloppypar}
\vspace{2mm}
\begin{sloppypar}
\noindent
Z.~Ajaltouni,
F.~Badaud,
G.~Chazelle,
O.~Deschamps,
S.~Dessagne,
A.~Falvard,
P.~Gay,
C.~Guicheney,
P.~Henrard,
J.~Jousset,
B.~Michel,
S.~Monteil,
\mbox{J-C.~Montret},
D.~Pallin,
J.M.~Pascolo,
P.~Perret,
F.~Podlyski
\nopagebreak
\begin{center}
\parbox{15.5cm}{\sl\samepage
Laboratoire de Physique Corpusculaire, Universit\'e Blaise Pascal,
IN$^{2}$P$^{3}$-CNRS, Clermont-Ferrand, F-63177 Aubi\`{e}re, France}
\end{center}\end{sloppypar}
\vspace{2mm}
\begin{sloppypar}
\noindent
J.D.~Hansen,
J.R.~Hansen,
P.H.~Hansen,$^{1}$
B.S.~Nilsson,
A.~W\"a\"an\"anen
\nopagebreak
\begin{center}
\parbox{15.5cm}{\sl\samepage
Niels Bohr Institute, 2100 Copenhagen, DK-Denmark$^{9}$}
\end{center}\end{sloppypar}
\vspace{2mm}
\begin{sloppypar}
\noindent
G.~Daskalakis,
A.~Kyriakis,
C.~Markou,
E.~Simopoulou,
A.~Vayaki
\nopagebreak
\begin{center}
\parbox{15.5cm}{\sl\samepage
Nuclear Research Center Demokritos (NRCD), GR-15310 Attiki, Greece}
\end{center}\end{sloppypar}
\vspace{2mm}
\begin{sloppypar}
\noindent
A.~Blondel,$^{12}$
\mbox{J.-C.~Brient},
F.~Machefert,
A.~Roug\'{e},
M.~Swynghedauw,
R.~Tanaka
\linebreak
H.~Videau
\nopagebreak
\begin{center}
\parbox{15.5cm}{\sl\samepage
Laboratoire de Physique Nucl\'eaire et des Hautes Energies, Ecole
Polytechnique, IN$^{2}$P$^{3}$-CNRS, \mbox{F-91128} Palaiseau Cedex, France}
\end{center}\end{sloppypar}
\vspace{2mm}
\begin{sloppypar}
\noindent
E.~Focardi,
G.~Parrini,
K.~Zachariadou
\nopagebreak
\begin{center}
\parbox{15.5cm}{\sl\samepage
Dipartimento di Fisica, Universit\`a di Firenze, INFN Sezione di Firenze,
I-50125 Firenze, Italy}
\end{center}\end{sloppypar}
\vspace{2mm}
\begin{sloppypar}
\noindent
A.~Antonelli,
M.~Antonelli,
G.~Bencivenni,
G.~Bologna,$^{4}$
F.~Bossi,
P.~Campana,
G.~Capon,
V.~Chiarella,
P.~Laurelli,
G.~Mannocchi,$^{5}$
F.~Murtas,
G.P.~Murtas,
L.~Passalacqua,
M.~Pepe-Altarelli
\nopagebreak
\begin{center}
\parbox{15.5cm}{\sl\samepage
Laboratori Nazionali dell'INFN (LNF-INFN), I-00044 Frascati, Italy}
\end{center}\end{sloppypar}
\vspace{2mm}
\begin{sloppypar}
\noindent
M.~Chalmers,
J.~Kennedy,
J.G.~Lynch,
P.~Negus,
V.~O'Shea,
B.~Raeven,
D.~Smith,
P.~Teixeira-Dias,
A.S.~Thompson
\nopagebreak
\begin{center}
\parbox{15.5cm}{\sl\samepage
Department of Physics and Astronomy, University of Glasgow, Glasgow G12
8QQ,United Kingdom$^{10}$}
\end{center}\end{sloppypar}
\begin{sloppypar}
\noindent
R.~Cavanaugh,
S.~Dhamotharan,
C.~Geweniger,$^{1}$
P.~Hanke,
V.~Hepp,
E.E.~Kluge,
G.~Leibenguth,
A.~Putzer,
K.~Tittel,
S.~Werner,$^{19}$
M.~Wunsch$^{19}$
\nopagebreak
\begin{center}
\parbox{15.5cm}{\sl\samepage
Kirchhoff-Institut f\"ur Physik, Universit\"at Heidelberg, D-69120
Heidelberg, Germany$^{16}$}
\end{center}\end{sloppypar}
\vspace{2mm}
\begin{sloppypar}
\noindent
R.~Beuselinck,
D.M.~Binnie,
W.~Cameron,
G.~Davies,
P.J.~Dornan,
M.~Girone,
N.~Marinelli,
J.~Nowell,
H.~Przysiezniak,$^{1}$
J.K.~Sedgbeer,
J.C.~Thompson,$^{14}$
E.~Thomson,$^{23}$
R.~White
\nopagebreak
\begin{center}
\parbox{15.5cm}{\sl\samepage
Department of Physics, Imperial College, London SW7 2BZ,
United Kingdom$^{10}$}
\end{center}\end{sloppypar}
\vspace{2mm}
\begin{sloppypar}
\noindent
V.M.~Ghete,
P.~Girtler,
E.~Kneringer,
D.~Kuhn,
G.~Rudolph
\nopagebreak
\begin{center}
\parbox{15.5cm}{\sl\samepage
Institut f\"ur Experimentalphysik, Universit\"at Innsbruck, A-6020
Innsbruck, Austria$^{18}$}
\end{center}\end{sloppypar}
\vspace{2mm}
\begin{sloppypar}
\noindent
C.K.~Bowdery,
P.G.~Buck,
D.P.~Clarke,
G.~Ellis,
A.J.~Finch,
F.~Foster,
G.~Hughes,
R.W.L.~Jones,
N.A.~Robertson,
M.~Smizanska
\nopagebreak
\begin{center}
\parbox{15.5cm}{\sl\samepage
Department of Physics, University of Lancaster, Lancaster LA1 4YB,
United Kingdom$^{10}$}
\end{center}\end{sloppypar}
\vspace{2mm}
\begin{sloppypar}
\noindent
I.~Giehl,
F.~H\"olldorfer,
K.~Jakobs,
K.~Kleinknecht,
M.~Kr\"ocker,
A.-S.~M\"uller,
H.-A.~N\"urnberger,
G.~Quast,$^{1}$
B.~Renk,
E.~Rohne,
H.-G.~Sander,
S.~Schmeling,
H.~Wachsmuth,
C.~Zeitnitz,
T.~Ziegler
\nopagebreak
\begin{center}
\parbox{15.5cm}{\sl\samepage
Institut f\"ur Physik, Universit\"at Mainz, D-55099 Mainz, Germany$^{16}$}
\end{center}\end{sloppypar}
\vspace{2mm}
\begin{sloppypar}
\noindent
A.~Bonissent,
J.~Carr,
P.~Coyle,
C.~Curtil,
A.~Ealet,
D.~Fouchez,
O.~Leroy,
T.~Kachelhoffer,
P.~Payre,
D.~Rousseau,
A.~Tilquin
\nopagebreak
\begin{center}
\parbox{15.5cm}{\sl\samepage
Centre de Physique des Particules de Marseille, Univ M\'editerran\'ee,
IN$^{2}$P$^{3}$-CNRS, F-13288 Marseille, France}
\end{center}\end{sloppypar}
\vspace{2mm}
\begin{sloppypar}
\noindent
M.~Aleppo,
S.~Gilardoni,
F.~Ragusa
\nopagebreak
\begin{center}
\parbox{15.5cm}{\sl\samepage
Dipartimento di Fisica, Universit\`a di Milano e INFN Sezione di
Milano, I-20133 Milano, Italy.}
\end{center}\end{sloppypar}
\vspace{2mm}
\begin{sloppypar}
\noindent
H.~Dietl,
G.~Ganis,
A.~Heister,
K.~H\"uttmann,
G.~L\"utjens,
C.~Mannert,
W.~M\"anner,
\mbox{H.-G.~Moser},
S.~Schael,
R.~Settles,$^{1}$
H.~Stenzel,
W.~Wiedenmann,
G.~Wolf
\nopagebreak
\begin{center}
\parbox{15.5cm}{\sl\samepage
Max-Planck-Institut f\"ur Physik, Werner-Heisenberg-Institut,
D-80805 M\"unchen, Germany\footnotemark[16]}
\end{center}\end{sloppypar}
\vspace{2mm}
\begin{sloppypar}
\noindent
P.~Azzurri,
J.~Boucrot,$^{1}$
O.~Callot,
M.~Davier,
L.~Duflot,
\mbox{J.-F.~Grivaz},
Ph.~Heusse,
A.~Jacholkowska,$^{1}$
L.~Serin,
\mbox{J.-J.~Veillet},
I.~Videau,$^{1}$
J.-B.~de~Vivie~de~R\'egie,
D.~Zerwas
\nopagebreak
\begin{center}
\parbox{15.5cm}{\sl\samepage
Laboratoire de l'Acc\'el\'erateur Lin\'eaire, Universit\'e de Paris-Sud,
IN$^{2}$P$^{3}$-CNRS, F-91898 Orsay Cedex, France}
\end{center}\end{sloppypar}
\vspace{2mm}
\begin{sloppypar}
\noindent
G.~Bagliesi,
T.~Boccali,
G.~Calderini,
V.~Ciulli,
L.~Fo\`a,
A.~Giammanco,
A.~Giassi,
F.~Ligabue,
A.~Messineo,
F.~Palla,$^{1}$
G.~Rizzo,
G.~Sanguinetti,
A.~Sciab\`a,
G.~Sguazzoni,
R.~Tenchini,$^{1}$
A.~Venturi,
P.G.~Verdini
\samepage
\begin{center}
\parbox{15.5cm}{\sl\samepage
Dipartimento di Fisica dell'Universit\`a, INFN Sezione di Pisa,
e Scuola Normale Superiore, I-56010 Pisa, Italy}
\end{center}\end{sloppypar}
\vspace{2mm}
\begin{sloppypar}
\noindent
G.A.~Blair,
J.~Coles,
G.~Cowan,
M.G.~Green,
D.E.~Hutchcroft,
L.T.~Jones,
T.~Medcalf,
J.A.~Strong,
\mbox{J.H.~von~Wimmersperg-Toeller} 
\nopagebreak
\begin{center}
\parbox{15.5cm}{\sl\samepage
Department of Physics, Royal Holloway \& Bedford New College,
University of London, Surrey TW20 OEX, United Kingdom$^{10}$}
\end{center}\end{sloppypar}
\vspace{2mm}
\begin{sloppypar}
\noindent
R.W.~Clifft,
T.R.~Edgecock,
P.R.~Norton,
I.R.~Tomalin
\nopagebreak
\begin{center}
\parbox{15.5cm}{\sl\samepage
Particle Physics Dept., Rutherford Appleton Laboratory,
Chilton, Didcot, Oxon OX11 OQX, United Kingdom$^{10}$}
\end{center}\end{sloppypar}
\vspace{2mm}
\begin{sloppypar}
\noindent
\mbox{B.~Bloch-Devaux},
D.~Boumediene,
P.~Colas,
B.~Fabbro,
G.~Fa\"{\i}f,
E.~Lan\c{c}on,
\mbox{M.-C.~Lemaire},
E.~Locci,
P.~Perez,
J.~Rander,
\mbox{J.-F.~Renardy},
A.~Rosowsky,
P.~Seager,$^{13}$
A.~Trabelsi,$^{21}$
B.~Tuchming,
B.~Vallage
\nopagebreak
\begin{center}
\parbox{15.5cm}{\sl\samepage
CEA, DAPNIA/Service de Physique des Particules,
CE-Saclay, F-91191 Gif-sur-Yvette Cedex, France$^{17}$}
\end{center}\end{sloppypar}
\vspace{2mm}
\begin{sloppypar}
\noindent
S.N.~Black,
J.H.~Dann,
C.~Loomis,
H.Y.~Kim,
N.~Konstantinidis,
A.M.~Litke,
M.A. McNeil,
G.~Taylor
\nopagebreak
\begin{center}
\parbox{15.5cm}{\sl\samepage
Institute for Particle Physics, University of California at
Santa Cruz, Santa Cruz, CA 95064, USA$^{22}$}
\end{center}\end{sloppypar}
\vspace{2mm}
\begin{sloppypar}
\noindent
C.N.~Booth,
S.~Cartwright,
F.~Combley,
P.N.~Hodgson,
M.~Lehto,
L.F.~Thompson
\nopagebreak
\begin{center}
\parbox{15.5cm}{\sl\samepage
Department of Physics, University of Sheffield, Sheffield S3 7RH,
United Kingdom$^{10}$}
\end{center}\end{sloppypar}
\vspace{2mm}
\begin{sloppypar}
\noindent
K.~Affholderbach,
A.~B\"ohrer,
S.~Brandt,
C.~Grupen,$^{1}$
J.~Hess,
A.~Misiejuk,
G.~Prange,
U.~Sieler
\nopagebreak
\begin{center}
\parbox{15.5cm}{\sl\samepage
Fachbereich Physik, Universit\"at Siegen, D-57068 Siegen, Germany$^{16}$}
\end{center}\end{sloppypar}
\vspace{2mm}
\begin{sloppypar}
\noindent
C.~Borean,
G.~Giannini,
B.~Gobbo
\nopagebreak
\begin{center}
\parbox{15.5cm}{\sl\samepage
Dipartimento di Fisica, Universit\`a di Trieste e INFN Sezione di Trieste,
I-34127 Trieste, Italy}
\end{center}\end{sloppypar}
\vspace{2mm}
\begin{sloppypar}
\noindent
H.~He,
J.~Putz,
J.~Rothberg,
S.~Wasserbaech
\nopagebreak
\begin{center}
\parbox{15.5cm}{\sl\samepage
Experimental Elementary Particle Physics, University of Washington, Seattle,
WA 98195 U.S.A.}
\end{center}\end{sloppypar}
\vspace{2mm}
\begin{sloppypar}
\noindent
S.R.~Armstrong,
K.~Cranmer,
P.~Elmer,
D.P.S.~Ferguson,
Y.~Gao,
S.~Gonz\'{a}lez,
O.J.~Hayes,
H.~Hu,
S.~Jin,
J.~Kile,
P.A.~McNamara III,
J.~Nielsen,
W.~Orejudos,
Y.B.~Pan,
Y.~Saadi,
I.J.~Scott,
J.~Walsh,
J.~Wu,
Sau~Lan~Wu,
X.~Wu,
G.~Zobernig
\nopagebreak
\begin{center}
\parbox{15.5cm}{\sl\samepage
Department of Physics, University of Wisconsin, Madison, WI 53706,
USA$^{11}$}
\end{center}\end{sloppypar}
}
\footnotetext[1]{Also at CERN, 1211 Geneva 23, Switzerland.}
\footnotetext[2]{Now at Universit\'e de Lausanne, 1015 Lausanne, Switzerland.}
\footnotetext[3]{Also at Dipartimento di Fisica di Catania and INFN Sezione di
 Catania, 95129 Catania, Italy.}
\footnotetext[4]{Also Istituto di Fisica Generale, Universit\`{a} di
Torino, 10125 Torino, Italy.}
\footnotetext[5]{Also Istituto di Cosmo-Geofisica del C.N.R., Torino,
Italy.}
\footnotetext[6]{Supported by the Commission of the European Communities,
contract ERBFMBICT982894.}
\footnotetext[7]{Supported by CICYT, Spain.}
\footnotetext[8]{Supported by the National Science Foundation of China.}
\footnotetext[9]{Supported by the Danish Natural Science Research Council.}
\footnotetext[10]{Supported by the UK Particle Physics and Astronomy Research
Council.}
\footnotetext[11]{Supported by the US Department of Energy, grant
DE-FG0295-ER40896.}
\footnotetext[12]{Now at Departement de Physique Corpusculaire, Universit\'e de
Gen\`eve, 1211 Gen\`eve 4, Switzerland.}
\footnotetext[13]{Supported by the Commission of the European Communities,
contract ERBFMBICT982874.}
\footnotetext[14]{Also at Rutherford Appleton Laboratory, Chilton, Didcot, UK.}
\footnotetext[15]{Permanent address: Universitat de Barcelona, 08208 Barcelona,
Spain.}
\footnotetext[16]{Supported by the Bundesministerium f\"ur Bildung,
Wissenschaft, Forschung und Technologie, Germany.}
\footnotetext[17]{Supported by the Direction des Sciences de la
Mati\`ere, C.E.A.}
\footnotetext[18]{Supported by the Austrian Ministry for Science and Transport.}
\footnotetext[19]{Now at SAP AG, 69185 Walldorf, Germany}
\footnotetext[20]{Now at Harvard University, Cambridge, MA 02138, U.S.A.}
\footnotetext[21]{Now at D\'epartement de Physique, Facult\'e des Sciences de Tunis, 1060 Le Belv\'ed\`ere, Tunisia.}
\footnotetext[22]{Supported by the US Department of Energy,
grant DE-FG03-92ER40689.}
\footnotetext[23]{Now at Department of Physics, Ohio State University, Columbus, OH 43210-1106, U.S.A.}
%
\setlength{\parskip}{\saveparskip}
\setlength{\textheight}{\savetextheight}
\setlength{\topmargin}{\savetopmargin}
\setlength{\textwidth}{\savetextwidth}
\setlength{\oddsidemargin}{\saveoddsidemargin}
\setlength{\topsep}{\savetopsep}
\normalsize
\newpage
\pagestyle{plain}
\setcounter{page}{1}

\section{Introduction}

  In general, neutral Higgs bosons do not couple directly to massless photons.
  For instance the standard model Higgs boson couples to photons 
  only through loops of charged particles, {\it i.e.,} W's, quarks 
  and leptons, and the branching ratio into $\gamma\gamma$ is 
  small ($\approx 10^{-3}$~\cite{kn:th0} for 
  $\mathrm{m_{H}} \sim 90$\,GeV/$\mathrm{c^2}$). However
  the $\gamma\gamma$ branching fraction of Higgs bosons can be increased 
  with respect to the standard model prediction in each of the following
  four configurations.

\begin{itemize}
\item The direct couplings to fermions are suppressed, as is the case
      for models with at least two Higgs multiplets~\cite{kn:th3}, of which one 
      couples only to fermions and the others only to gauge bosons. 
      The physical states with couplings only to gauge bosons are called 
      {\it fermiophobic} Higgs bosons.

\item The direct couplings to gauge bosons are enhanced with anomalous
      couplings~ \cite{kn:th2}. These couplings are described in the most general 
      formulation with four effective six-dimensional operators
      with strength $\mathrm{f_{\it{i}}/\Lambda^2}$, where $\Lambda$ is the scale of 
      the new underlying interaction.
\item Couplings to both fermions and bosons are modified, as is the 
      case in the minimal supersymmetric extension of the standard 
      model (MSSM).

\item Additional light, charged particles enter the loops that 
      couple Higgs bosons and photons, as is again the case in the 
      MSSM (charginos, squarks, sleptons, charged Higgs bosons).
\end{itemize}
   With some particular choices of parameters, the branching ratio
   into $\gamma\gamma$ may be enhanced in the 
   MSSM, and can reach a value close to $100\%$ in models with 
   fermiophobia or with anomalous gauge couplings. It is therefore 
   possible that a Higgs boson has escaped the standard search for the 
   Higgs-strahlung process ${\mathrm e}^+{\mathrm e}^- \to {\mathrm H}
   {\mathrm f}\bar{\mathrm f}$ with ${\mathrm H} \to {\mathrm b}
   \bar{\mathrm b}$~\cite{kn:aleph1}.
   In this letter, a complementary search for the Higgs-strahlung 
   process with ${\mathrm H} \to \gamma\gamma$ is described.

   The analysis addresses all topologies arising from the ${\mathrm e}^+
   {\mathrm e}^- \to {\mathrm HZ^{(*)}}$ process, 
   as characterized by the charged track multiplicity of the final state: {\it (i)}
   acoplanar photons with missing energy and  no charged particles
   for ${\mathrm H}\nu\bar\nu$; {\it (ii)} photon pairs with exactly two
   charged particles identified as leptons for $\mathrm{ H \ell^+\ell^-}$; 
   {\it (iii)} photon pairs accompanied with two thin, low multiplicity 
   jets for ${\mathrm H}\tau^+\tau^-$, from two to four charged particles; 
   and {\it (iv)} photon pairs with a hadronic system for ${\mathrm H}
   {\mathrm q}\bar{\mathrm q}$ with at least five charged particles.

The analysis is performed with the data collected with the ALEPH detector
from 1991 to 1999 including the Z peak data collected during the
LEP\,2 period. This sample corresponds to an integrated luminosity of 
$672\,\mathrm{pb}^{-1}$ at centre-of-mass energies ranging 
from 88 to 202~\,GeV. Details are given in Table~\ref{tab:lumi}.

\begin{table}[htb]
\caption{\protect\footnotesize Integrated luminosity for the Z peak data (88-94\,GeV) 
and high energy data (130-202\,GeV). The energies are rounded to the closest
integer value.}
\label{tab:lumi}
\centering
\begin{tabular}{||l||c|c|c|c|c|c|c|c|c||} \hline
$\mathrm{\sqrt{\it s}}$ (GeV) & 88 & 89 & 90 & 91 & 92 & 93 & 94 & 130 & 136 \\ \hline
L ($\mathrm{pb^{-1}}$) & 0.7 & 17.9 & 0.8 & 124.8 & 0.8 & 19.1 & 0.8 & 6.2 & 6.4 \\ \hline \hline
$\mathrm{\sqrt{\it s}}$ (GeV) & 161 & 170 & 172 & 183 & 189 & 192 & 196 & 200 & 202  \\ \hline
L ($\mathrm{pb^{-1}}$) & 11.1 & 1.1 & 9.5 & 59.2 & 177.1 & 28.9 & 79.8 & 86.3 & 42.0  \\ \hline
\end{tabular}
\end{table}

   After a short description of the detector properties relevant for
   this search, the common preselection based on photon identification 
   is reviewed in Section 3. The global search strategy is developed
   in Section 4. The systematic uncertainties affecting the 
   selection efficiency are discussed in Section 5 and the results
   are given in Section 6.

\section{The ALEPH detector}

The ALEPH detector and its performance are described in Refs. \cite{kn:ref1,kn:ref2}.
The tracking detectors, composed
of the silicon vertex detector surrounded by the
inner tracking chamber and the time projection chamber (TPC), provide efficient 
reconstruction of charged particles in the angular range $|{\cos\theta}|<0.96$. 
A charged particle track is called
a {\em good track} if it is reconstructed with a least 
four hits in the TPC and if it originated from within a cylinder of length 
20\,cm and radius 2\,cm, coaxial with the beam and centred at the interaction point. 
A 1.5\,T axial
magnetic field delivered by a super-conducting solenoidal coil allows a
charged particle $\mathrm{1/p_{\perp}}$ resolution of 
$\mathrm{(6\times 10^{-4}\bigoplus 5 \times 10^{-3}/p_{\perp})(GeV/{\it c})^{-1}}$ to be
achieved.

The electromagnetic calorimeter (ECAL) is a lead/wire-plane
sampling calorimeter covering the angular range $|{\cos\theta}|< 0.98$.
Anode wire signals provide a measurement of the arrival time
of the particles relative to the beam crossing 
with a resolution better than 15\,ns.
Cathode pads associated with each wire layer
are connected to form projective towers of approximately
$0.9^{\circ}$ by $0.9^{\circ}$
which are read out in three segments in depth.
The impact parameter of the photon with respect to 
the interaction point is estimated from the 
barycentre of the electromagnetic shower in each segment
with a resolution of about 6\,cm.
A photon candidate is identified using a topological search~\cite{kn:ref2}
for energy deposits in neighbouring electromagnetic calorimeter towers
isolated from the extrapolation of any charged particle track to the ECAL.
Any photon candidate close to a boundary between ECAL modules
or pointing towards an 
uninstrumented region of the TPC is not considered in the analysis. 
The energy calibration of the ECAL is obtained from Bhabha events, radiative
returns to the Z resonance, $\mathrm{e^+e^-\to\gamma\gamma}$ and
$\mathrm{\gamma\gamma\to e^+e^-}$ events.
The energy resolution for photons is
$\mathrm{\delta E/E = 0.25/\sqrt{E/GeV}+0.009}$ \cite{kn:ref2}.

The luminosity monitors (LCAL and SICAL) extend the calorimetric coverage down to
small polar angles.
The iron return yoke is instrumented with streamer tubes and acts as a hadron
calorimeter (HCAL), covering polar angles down to 110\,mrad. Surrounding the
HCAL are two additional double layers of streamer tubes called muon chambers.

The measurements of the tracking detectors and the calorimeters
are combined into objects classified as charged particles, photons and neutral hadrons using
the energy flow algorithm described in Ref.\cite{kn:ref2}. All objects are
used to compute the total visible energy $\mathrm{E_{vis}}$ and the
missing energy $\mathrm{E_{miss}}$ with a resolution of 
($\mathrm{0.6 \sqrt{E_{vis}/GeV}+0.6}$) GeV.
 
Electron identification is based on
the matching between the measured momentum in the tracking system
and the energy in the ECAL,
the shower profile in the ECAL 
and the measurement of the specific ionisation energy loss in the TPC.
Muons are identified by their characteristic hit patterns in the HCAL
and in the muon chambers.

\section{Event preselection}

Signal events are characterized by two isolated, energetic photons,
well contained in the
apparatus, and in time with the beam crossing.

Photon isolation is ensured by requiring that the total charged energy in a
cone of half-angle $14^\circ$ around the photon direction ($\mathrm{E_{14^{\circ}}^\gamma}$) 
be smaller than 2\,GeV, and that the
invariant mass between the photon and any charged particle
($\mathrm{m}_{\gamma,\mathrm{ch}}$) be in excess of 1\,$\mathrm{GeV/{\it c}^2}$.
Pairs of photons from a $\pi^0$ decay are rejected by the requirement
that their invariant mass be larger than $\mathrm{1\,GeV/{\it c}^2}$.

Photon centrality is enforced by requiring that the polar angles satisfy 
$|{\cos\theta_\gamma}|< 0.9$. Since the Higgs boson is expected to be
produced with a nearly uniform $\cos\theta$ distribution,
the sum of the two cosines (in absolute value) must be
smaller than 1.4, and the cosines of the Higgs boson production
and the thrust axis polar angles, $\mathrm{\theta_{\gamma\gamma}}$ and
$\mathrm{\theta_{thrust}}$ must be within $\pm 0.95$. Events with energy at low polar
angles are further vetoed by rejecting
events with more than 2\,GeV within $14^\circ$ of the
beam axis ($\mathrm{E_{14^{\circ}}^{beam}})$.

If, in a selected event, more than two photons satisfy the above criteria, only the
most energetic two photons are considered as originating from a Higgs boson
candidate.

Finally, only events in time with the beam crossing are kept. For events with
at least two good tracks, a timing to better than 1\,ns is ensured by the good track
definition. Events with only one good track are rejected. The production time of events
with no good tracks is determined as the energy-weighted average of the times
reconstructed in all ECAL modules, t$_0$, and is required to be in agreement with the beam
crossing time within $\pm 40$\,ns. For these events, the impact parameter of each
identified photon is also required to be less than 25\,cm.

The reconstructed particles of events with at least two good tracks are then forced
to form four ``jets'' with the Durham jet clustering algorithm~\cite{kn:duhram}.
The consistency of these events
with a four-body final-state hypothesis is verified by requiring that each of the four jet
energies $\mathrm{E_{\it{i}}^{resc}}$, rescaled to satisfy energy-momentum
conservation under the assumption that the jet velocities are perfectly measured, be
positive. However, in order to make the reconstructed Higgs boson mass
resolution independent of the final state topology, the measured photon
energies were used instead of the rescaled energies.

In the following, the {\it neutral electromagnetic energy} in a given jet is computed
with all neutral objects in the electromagnetic calorimeter, and with the objects in the
hadron calorimeter found behind uninstrumented regions of the electromagnetic calorimeter.
An {\it electromagnetic jet} is defined as a jet with more than $80\%$ of electromagnetic
energy. In events reconstructed with four jets, {\it i.e.}, with at least two good tracks,
the two jets with the smallest electromagnetic energy fraction are called {\it fermionic
jets}.

At this level, a signal efficiency of 20 to $65\%$ is achieved at all centre-of-mass
energies and for any Higgs boson mass above 1\,$\mathrm{GeV/{\it c}^2}$,
as estimated with many ${\mathrm e}^+{\mathrm e}^- \to {\mathrm{Hf}}\bar{\mathrm f}$
simulated event samples produced with
the {\tt HZHA} generator~\cite{kn:dmc1} and processed through
the whole detector simulation and event reconstruction chain.

\section{Event selection}

Events passing the above preselection are further
classified in the four signal
topologies according to their good track multiplicity ($\mathrm{n_{ch}}$).

\begin{itemize}

\item Events with no good tracks are classified as $\mathrm{H\nu\bar{\nu}}$ candidate
events. The main standard model background sources to this final state
are {\it(i)}
$\mathrm{e}^{+}\mathrm{e}^{-}\rightarrow\nu\bar{\nu}\gamma(\gamma)$,
simulated with the {\tt KORALZ} package~\cite{kn:dmc3};
and {\it (ii)} $\mathrm{e^+e^-\to\gamma\gamma(\gamma)}$, simulated with
the {\tt GGG} generator~\cite{kn:dmc2}. 
The latter generator does not contain QED
contributions of order $\alpha^4$ and above, but they were
estimated in Ref.~\cite{kn:gary} to be small enough to be considered 
negligible in the present analysis.

\item Events with two good tracks, both positively
identified as electrons or muons, are classified as $\mathrm{H\ell^+\ell^-}$
(with $\ell$ = $\mathrm{e}$ or $\mu$) candidate events.
The programs {\tt BHWIDE}~\cite{kn:dmc4} and {\tt UNIBAB}~\cite{kn:dmc5} on
the one hand, and {\tt KORALZ} on the other, are employed to
simulate the main background processes,
{\it i.e.}, $\mathrm{e}^{+}\mathrm{e}^{-}\rightarrow\mathrm{e}^{+}\mathrm{e}^{-}\gamma(\gamma)$
and $\mathrm{\mu^{+}\mu^{-}\gamma(\gamma)}$, respectively.

\item Other two good track events, and events with up to four good tracks
are classified as $\mathrm{H}\tau^{+}\tau^{-}$ candidate events. Here again, the main
background process,
$\mathrm{e}^{+}\mathrm{e}^{-}\rightarrow\tau^{+}\tau^{-}\gamma(\gamma)$
is simulated with {\tt KORALZ}.

\item Finally, events with at least five good tracks are classified as
$\mathrm{Hq\bar{q}}$ candidate events.
The $\mathrm{e}^{+}\mathrm{e}^{-}\rightarrow\mathrm{q\bar{q}}(\gamma)$
process is simulated with {\tt JETSET}~\cite{kn:lund}  for data taken at the Z resonance,
and with {\tt PYTHIA}~\cite{kn:dmc6} and {\tt HERWIG}~\cite{kn:dmc7} at higher energies.        
The four-fermion processes WW, ZZ, Zee and $\mathrm{We}\nu$
are simulated using {\tt PYTHIA}~\cite{kn:dmc6}.

\end{itemize}

The selection criteria designed to reduce the contribution of the
background processes are summarized in Table~\ref{tab:cuts}, and only
a brief account is given here.

\begin{table}[!bt]
\caption{\protect\footnotesize
 Overview of all selection cuts}
\label{tab:cuts}
\centering
\vspace*{0.3cm}
\begin{tabular}{||c|c|c||} \hline
\multicolumn{2}{||c|}{\bf{Z peak}} & \bf{High energy} \\ \hline \hline
\multicolumn{2}{||c|}{${\mathrm{E_{\gamma_{\it{i}}}> 3\,GeV}}$;
            ${\mathrm{E}_{\gamma}^{max}> 0.1\sqrt{\mathrm{\it{s}}}}$}
 & ${0.2\mathrm{E_{beam}}<\mathrm{E_{\gamma_{\it{i}}}}< 0.75 \mathrm{E_{beam}}}$\\
\multicolumn{2}{||c|}{${\mathrm{E_H^{recoil}> 10\,GeV };
        \mathrm{E_{visible}}> 0.6\sqrt{\mathrm{\it{s}}}}$}
 & ${|{\mathrm{ m_{rec}-m_{Z}}}|< 15\,\mathrm{GeV}/{\it c}^2}$\\ \hline
\multicolumn{3}{||c||}{\em{No good track}:{$\,\mathrm{H\nu\bar\nu}$}}\\ \hline
\multicolumn{3}{||c||}{$\,\,\,\,\,\,\,\,\,\,\,\,\,\,\,\,\,\,\,\,\,\,\,\,\,\,\,\,\,\,
|{\cos\theta_{\mathrm{\gamma\gamma}}}|< 0.9$}\\
\multicolumn{2}{||c|}{$\mathrm{E}_{miss}> 15\,\mathrm{GeV}$} & \\
\multicolumn{2}{||c|}{${\mathrm{E_{vis}-E_{\gamma\gamma}}< 10\, \mathrm{GeV}}$}
  & ${\mathrm{E}_{\gamma}^{max}> 0.15\sqrt{\mathrm{\it{s}}}}$ \\
\multicolumn{2}{||c|}{$\mathrm{|{m_{\gamma\gamma}+m_{rec}-\sqrt{\it s}}|>
2\,GeV/{\it c}^2}$}
  &\\  
\multicolumn{2}{||c|}{${|{\cos\theta_{\gamma\gamma}^{\mathrm{dcy}}}|< 0.95}$} &\\ \hline
\multicolumn{3}{||c||}{\em{Two good tracks}:{$\,\mathrm{H\ell^+\ell^-/H\tau^+\tau^-}$}}\\ \hline
\multicolumn{3}{||c||}{$\,\,\,\,\,\,\,\,\,\,\,\,\,\,\,\,\,\,\,\,\,\,\,\,\,\,\,\,\,\,
\mathrm{E}_{\it{i}}^{\mathrm{resc}}> 0\,\mathrm{GeV}$}\\
\multicolumn{3}{||c||}{$\,\,\,\,\,\,\,\,\,\,\,\,\,\,\,\,\,\,\,\,\,\,\,\,\,\,\,\,\,\,
\mathrm{m_{f\bar{f}}}> 1\, \mathrm{GeV/{\it c}^{2}}$}\\
\multicolumn{3}{||c||}{$\,\,\,\,\,\,\,\,\,\,\,\,\,\,\,\,\,\,\,\,\,\,\,\,\,\,\,\,\,\,
\mathrm{m_{\gamma,ch}}> 5\, \mathrm{GeV/{\it c}^2}$}\\
\multicolumn{2}{||c|}{$\mathrm{p_{\perp_{ch}}^{\gamma}}> 2\,\mathrm{GeV/{\it c}}$} &\\
\multicolumn{2}{||c|}{$\mathrm{m_{\gamma,f}> 10\,GeV/{\it{c}}^2}$} 
  &   \\ \cline{1-2} 
$\mathrm{e^+e^-/\mu^+\mu^-:H\ell^+\ell^-}$ &
$\mathrm{e\mu/e h/\mu h/hh:H\tau^+\tau^-}$
& \,\,\,\,\,\,\,\,\,\,\,\,\,\,\,\,\,\,\,\,\,\,
$\mathrm{p_{\perp_{ch}}^{\gamma}}> 10\,\mathrm{GeV/\it{c}}$
\,\,\,\,\,\,\,\,\,\,\,\,\,\,\,\,\,\,\,\,\,\,\, \\ \cline{1-2}  
$\mathrm{E_{\gamma}> 10\,GeV}$
& $\mathrm{E_{\gamma}}> 20\,\mathrm{GeV}$
&  \\
$|{\cos\theta_{\ell^+\ell^-}^{\mathrm{dcy}}}|< 0.95$ 
& $|{\mathrm{m_{rec}-m_{f\bar{f}}^{resc}}}|< 20\,\mathrm{GeV/{\it c}^2}$ &\\ \hline
\multicolumn{3}{||c||}{{\em Three or four good tracks:}$\,\mathrm{H\tau^+\tau^-}$}\\ \hline
\multicolumn{3}{||c||}{$\,\,\,\,\,\,\,\,\,\,\,\,\,\,\,\,\,\,\,\,\,\,\,\,\,\,\,\,\,\,
\mathrm{E_{\it{i}}^{resc}> 0}\,\mathrm{GeV}$}\\
\multicolumn{2}{||c|}{$\mathrm{p_{\perp_{f}}^{thrust}}> 2\,
\mathrm{GeV/\it{c}}$;\,$\mathrm{m_{\gamma,f}}> 10\, \mathrm{GeV/{\it{c}}^2}$ } &\\
\multicolumn{2}{||c|}{$\mathrm{N_{jets}^{elec}}=2$;$\ \mathrm{E_{miss}}<
15\, \mathrm{GeV}$}
 & $\mathrm{p_{\perp_{f}}^{thrust}}> 4\, \mathrm{GeV/\it{c}}$  \\
\multicolumn{2}{||c|}{$|{\cos\theta_{\gamma\gamma}^{\mathrm{dcy}}}|< 0.95$;\, 
$|{\cos\theta_{\mathrm{f\bar{f}}}^{\mathrm{dcy}}}|< 0.95$} &\\ \hline
\multicolumn{3}{||c||}{{\em Five or more good tracks:}$\,\mathrm{Hq\bar{q}}$}\\ \hline
\multicolumn{3}{||c||}{$\,\,\,\,\,\,\,\,\,\,\,\,\,\,\,\,\,\,\,\,\,\,\,\,\,\,\,\,\,\,
\mathrm{E_{\it{i}}^{resc}> 0}\,\mathrm{GeV}$}\\
\multicolumn{3}{||c||}{$\,\,\,\,\,\,\,\,\,\,\,\,\,\,\,\,\,\,\,\,\,\,\,\,\,\,\,\,\,\,
|{\cos\theta_{\gamma}}|< 0.8$}\\
\multicolumn{3}{||c||}{$\,\,\,\,\,\,\,\,\,\,\,\,\,\,\,\,\,\,\,\,\,\,\,\,\,\,\,\,\,\,
\it{y}_{\mathrm{34}}\mathrm{> 0.001}$}\\
\multicolumn{2}{||c|}{$\mathrm{p_{\perp_{f}}^{thrust}}> 2\,
\mathrm{GeV/\it{c}}$;\,$\mathrm{m_{\gamma,f}}> 10\,
\mathrm{GeV/{\it{c}}^2}$ }
 &  $\mathrm{p_{\perp_{f}}^{thrust}}> 4\, \mathrm{GeV/\it{c}}$ \\
\multicolumn{2}{||c|}{$\mathrm{N_{jets}^{elec}}=2$;$\ \mathrm{E_{miss}}<
15\, \mathrm{GeV}$}
 &  $\mathrm{p_{\perp_{\gamma}}^{f}}> 0.05\sqrt{\mathrm{\it{s}}}$\\
\multicolumn{2}{||c|}{$|{\cos\theta_{\gamma\gamma}^{\mathrm{dcy}}}|< 0.95$;\, 
$|{\cos\theta_{\mathrm{f\bar{f}}}^{\mathrm{dcy}}}|< 0.95$}
 & $\mathrm{\theta_{4-jets}^{min}}> 350^\circ$\\ \hline \hline
\end{tabular}
\end{table}

For each of the topologies, the photons of the $\mathrm{e^+e^-\to f\bar{f}\gamma(\gamma)}$
background process are emitted by the incoming and the outgoing
charged fermions. They are therefore preferentially produced either
along the beam axis or along one of the outgoing charged fermion
directions. An efficient rejection is achieved by tightening
the photon isolation requirements with respect to these directions, by
means of the variables introduced in Section 3 and a number of other
relevant variables:
{\it(i)} photon energies $\mathrm{E_{\gamma}}$ and the energy recoiling against the photon
pair $\mathrm{E_H^{recoil}}$, {\it(ii)} invariant masses
($\mathrm{m_{\gamma,ch}}$, $\mathrm{m_{\gamma,f}}$) and transverse
momentum $\mathrm{p_{\perp_{ch}}^{\gamma}}$  
of individual charged particles and fermionic jets
with respect to the photons, {\it(iii)} transverse momentum of the photon
with respect to the closest fermionic jet $\mathrm{p_{\perp_{\gamma}}^{f}}$ and
transverse momentum of the fermionic jets with respect to the thrust 
axis $\mathrm{p_{\perp_{f}}^{thrust}}$, {\it(iv)} the value ${\it{y}}_{\mathrm{34}}$
of the Durham ${\it{y}}_{\mathrm{cut}}$ transition value between three and four
jets and the invariant mass of the two fermionic jets $\mathrm{m_{f\bar{f}}}$.
For the Z peak data, exactly two electomagnetic jets $\mathrm{N_{jets}^{elec}}$ are
required in order to
reject events in which both photons are inside the same jet.

Further cuts are made on the cosine of the decay angle of the photons
(fermions) in the rest frame of the $\gamma\gamma$ ($\mathrm{f\bar{f}}$)
system  $\cos\theta_{\gamma\gamma}^{\mathrm{dcy}}$
($\mathrm{\cos\theta_{f\bar{f}}^{dcy}}$). These decay angles are expected
to have a flat distribution for signal events, while they are strongly
peaked towards small angles for the background. For a high mass Higgs
at high energies the signal events are expected to be somewhat
spherical in nature. Non-spherical events are rejected
by a cut on the sum of the four minimum inter-jet angles 
$\mathrm{\theta_{4-jets}^{min}}$. At high energies the mass recoiling
against the photonic system $\mathrm{m_{rec}}$ is required to be consistent with
the Z mass. For the $\mathrm{H\tau\tau}$ channel at Z peak energies the mass
recoiling against the photonic system is required to be consistent with 
the rescaled mass of the fermionic system $\mathrm{m_{ff}^{resc}}$.
Finally, in the $\mathrm{H\nu\bar{\nu}}$ topology,
some boost is ensured by requiring that the total produced mass does not
saturate the available energy:
$\mathrm{|{m_{\gamma\gamma}+m_{rec}-\sqrt{\it s}}|>\,2\,GeV}$.

The typical efficiencies and expected standard
model background for each topology, together with
cross-channel contamination, are shown in Table~\ref{tab:summary}.

\begin{table}[htb]
\caption{\protect\footnotesize Efficiencies for the different Z decay channels
for a 110\,$\mathrm{GeV/{\it c}^2}$ Higgs boson mass at
$\mathrm{\sqrt{\it s}=202\,GeV}$ and numbers of expected background and
observed events at the Z peak (Z) and at high energies (HE).}
\label{tab:summary}
\centering
\begin{tabular}{||l||c|c|c|c||c|c|c|c||} \hline
$\mathrm{n_{ch}}$ & $\mathrm{\epsilon_{H\nu\bar{\nu}}}$
             & $\mathrm{\epsilon_{H\ell^+\ell^-}}$
             & $\mathrm{\epsilon_{H\tau^+\tau^-}}$
             & $\mathrm{\epsilon_{Hq\bar{q}}}$ 
             & $\mathrm{N_{exp.}^{back.}}$(Z)
             & $\mathrm{N_{obs.}}$(Z)
             & $\mathrm{N_{exp.}^{back.}}$(HE)
             & $\mathrm{N_{obs.}}$(HE) \\ \hline\hline
 0           & $47.0\%$ & $0.3\% $ & $0.0 $ & $0.0 $ & 0.9 &1 & 3.6 &0\\ \hline
 2           & $0.0 $ & $36.7\%$ & $22.4\%$ & $0.0 $ & 2.3 &4 & 2.9 &2\\ \hline
 3-4         & $0.0 $ & $2.0\% $ & $14.9\%$ & $0.0 $ & 0.7 &0 & 0.4 &2\\ \hline
 $\geq$5     & $0.0 $ & $0.0 $ & $5.2\% $ & $37.3\%$ & 9.3 &7 & 7.8 &6\\ \hline \hline
 Total       & $47.0\%$ & $39.0\%$ & $42.5\%$ & $37.3\%$ & 13.2&12& 14.7&10 \\ \hline \hline
\end{tabular}
\end{table}

\section{Systematic uncertainties}

Because no background subtraction is performed to derive the final result, the
uncertainty on the background evaluation has not been estimated.
The uncertainty on the photon selection efficiency
summarized in
Table~\ref{tab:systematic}
receives a large contribution from the photon
identification efficiency. The uncertainties due to photon energy calibration or
photon angle resolution affect only the $\gamma\gamma$ invariant mass
resolution.

\begin{figure}[!ht]
\centerline{\epsfxsize 14.5 truecm\epsfbox[0 90 567 567]{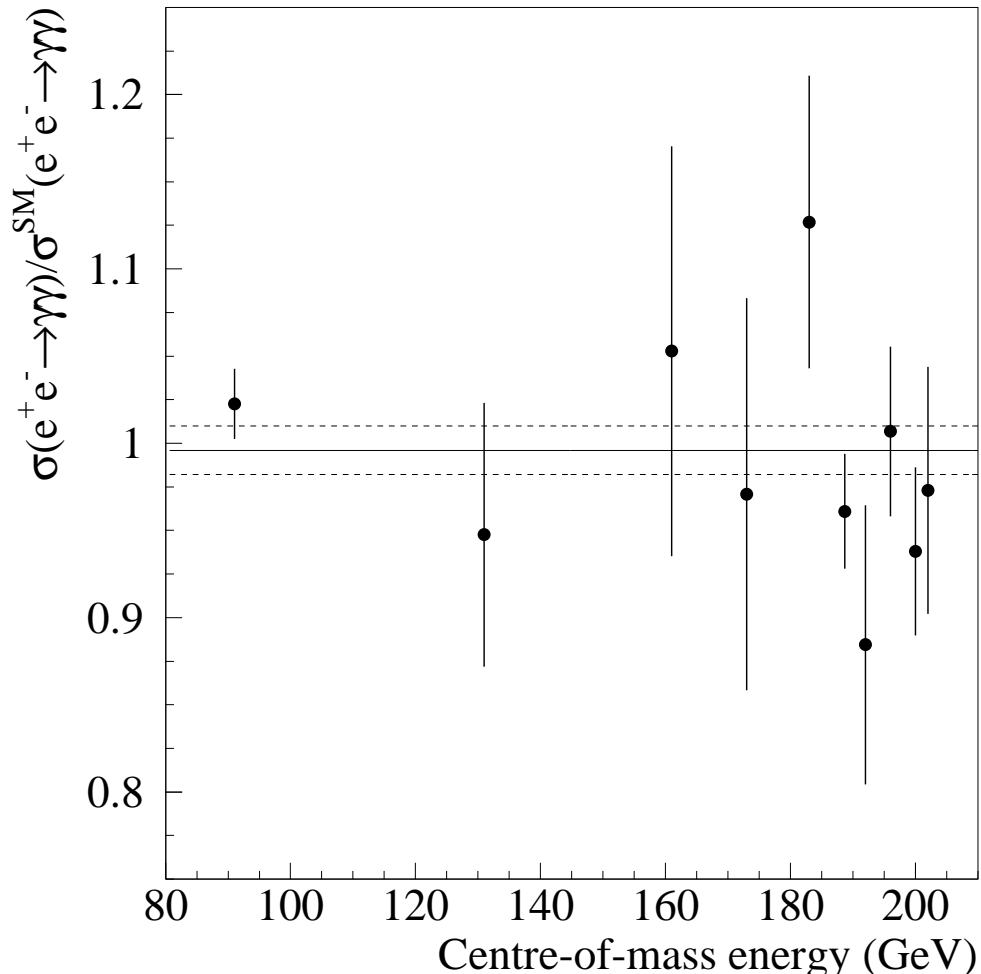}}
\vspace*{1.8cm}
\caption{\protect\footnotesize\label{fig:gggsect}{ Ratio of the
measured cross section of the
$e^{+}e^{-} \rightarrow \gamma\gamma(\gamma)$ process at all centre-of-mass
energies to the expected standard model cross section.
The solid line is the best fit to the data and the two dotted lines
represent the statistical error from the fit.}} 
\end{figure}

To estimate the uncertainty on the photon selection efficiency,
the total cross section of the 
$\mathrm{e^+e^- \to \gamma\gamma(\gamma)}$ process was measured and compared
to its prediction. Two-photon-candidate events are selected as final states
with only two identified photons (with the same timing and pointing
constraints as those defined in section 3)
with a polar angle such that $\mathrm{|{\cos{\theta_{\gamma}}}|< 0.9}$ ($0.95$), 
and an opening angle satisfying 
$\mathrm{\cos\alpha_{\gamma\gamma}< -0.999}$ ($-0.9999$), 
for the Z peak data (at high energy).
The $\mathrm{e^+e^-\to e^+e^-\gamma\gamma}$ background is rejected
by requiring no charged particles in
the event and not more than 2 GeV of energy around the beam direction
($\mathrm{E}_{14^{\circ}}^{\mathrm{beam}}< 2\,GeV$). The
$\nu\bar{\nu}\gamma\gamma$ background is reduced to a negligible amount by requiring
the invariant mass of the two-photon system to be greater than $0.75\sqrt{\it s}$.   
The result is displayed in Fig.~\ref{fig:gggsect}.
From this measurement, a conservative relative systematic uncertainty is 
estimated to be $3\%$, including a $1\%$ theoretical error.
The systematic uncertainty on the photon
detection efficiency includes the effects of the cuts
on photons used at preselection level as well as that of
the integrated luminosity determination, which enters the
$\mathrm{e^+e^- \to \gamma\gamma(\gamma)}$ cross section
measurement.  

The effect of the photon energy calibration and angular resolution on the efficiency is
estimated with Bhabha events. The relative difference between the values
of the electron energy determined independently by the tracking system
and by the calorimetry mainly originates from
electron Bremsstrahlung. This difference is measured to be $\sim$ $1.5\%$, and
it agrees within $0.2\%$ with that expected from the simulation. This $\pm 0.2\%$
uncertainty is conservatively assigned to the photon energy calibration.
It leads to a shift of the measured Higgs boson mass by 
$\pm 200$ $\mathrm{MeV/{\it c}^2}$, and to a negligible increase of the mass
resolution. The uncertainty on the reconstructed photon direction is estimated by
comparing the directions determined with the tracking and the calorimetry. It degrades the
measured Higgs boson mass resolution by approximately 20 $\mathrm{MeV/{\it c}^2}$ 
with a negligible loss of efficiency.
These systematic uncertainties are included in the final 
result by increasing the Higgs boson mass
resolution by $200\,\mathrm{MeV/{\it c}^2}$ .
\begin{table}[ht]
\caption{\protect\footnotesize Systematic uncertainties on signal efficiency}
\label{tab:systematic}
\centering
\begin{tabular}{||l||c||} \hline
\,\,\,\,\,\,\,\,\,\,\,\,\,\,\,\,\,\,\,\,\,Sources & Relative uncertainty in $\%$ \\ \hline
Photon selection efficiency           &  3.0   \\ \hline
$\gamma$ energy calibration           &  0.5 \\ \hline
$\gamma$ angular resolution           &  0.1 \\ \hline
Total energy calibration (Z peak data only)       &  0.5 \\ \hline  
Lepton identification                      &  0.1 \\ \hline
Photon isolation                           &  1.2 \\ \hline
$\cos\theta_{\mathrm{thrust}}$        &  0.1 \\ \hline
$\mathrm{p_{\perp_{f}}^{thrust}}$     &  0.2 \\ \hline
$\mathrm{m_{\gamma,f}}$               &  0.3 \\ \hline
$\mathrm{p_{\perp_{\gamma}}^{f}}$ (High energy data only) & 0.4 \\ \hline
$\mathrm{y_{34}}$                      &  0.5  \\ \hline
$\mathrm{\theta_{4-jets}^{min}}$ (High energy data only)      & 0.3 \\ \hline
Model dependence & 4.0 \\ \hline
Total in quadrature & 5.2 \\ \hline \hline
\end{tabular}
\end{table}

As the lepton identification is used only to classify events,
the signal efficiency decreases by less than $0.1\%$ relative 
when the lepton identification efficiency is modified by $10\%$.

The remaining selection criteria are expected to be largely insensitive
to the details of the simulation of the hadronic system. A quantitative
estimate of the size of possible discrepancies is performed with the aid
of an event reweighting technique. For each selection variable,
bin-by-bin correction factors are calculated as the ratio of data to
Monte Carlo expectation, evaluated at the preselection level with the
cut on $\mathrm{m_{\gamma,ch}}$ removed. The Monte Carlo signal 
distribution of the selection variable, obtained when all cuts are
applied, is then re-weighted with these correction factors to
obtain a new estimate of the efficiency. The difference in
efficiencies estimated with this technique are, for almost all
variables, at the level of a few parts per mil and are given in
Table~\ref{tab:systematic}. The largest effect, of $1.2\%$, comes from
the cut on the isolation of the photon $\mathrm{m_{\gamma,ch}}$.

Finally, the model dependence of this analysis is estimated with the anomalous coupling
model mentioned in Section 1. In that context, all the anomalous coupling parameters 
$\mathrm{{f_{\it{i}}/\Lambda^2}}$ are varied independently of each other within
$\mathrm{\pm 100}$ $\mathrm{TeV^{-2}}$ using the {\tt HZHA} generator. The influence of the
Higgs boson energy and angular distributions leads to a relative uncertainty
on the signal efficiency of at most $4\%$.

The decay width of the
Higgs boson is in general negligible with respect to the detector resolution
(below 1 $\mathrm{GeV}$ for ${\mathrm{|{f_{\it{i}}/\Lambda^2}|< 100}}$ $\mathrm{TeV^{-2}}$).
In anomalous coupling models with $\mathrm{|{f_{\it{i}}/\Lambda^2}|}$
above 500 $\mathrm{TeV^{-2}}$ 
the decay width of a heavy Higgs 
boson becomes larger than the detector invariant
mass resolution ($\mathrm{\Gamma}\approx 3$ $\mathrm{GeV}$) and the relative
variation of the signal efficiency becomes larger than the previously estimated systematic
uncertainties. The final results presented in the next section are therefore valid
for models in which the width of the Higgs boson is less than a few GeV.

\section{Results}

\begin{figure}[ht]
\vspace*{-1.cm}
{\epsfxsize 18.5 truecm\epsfbox[0 280 567 567]{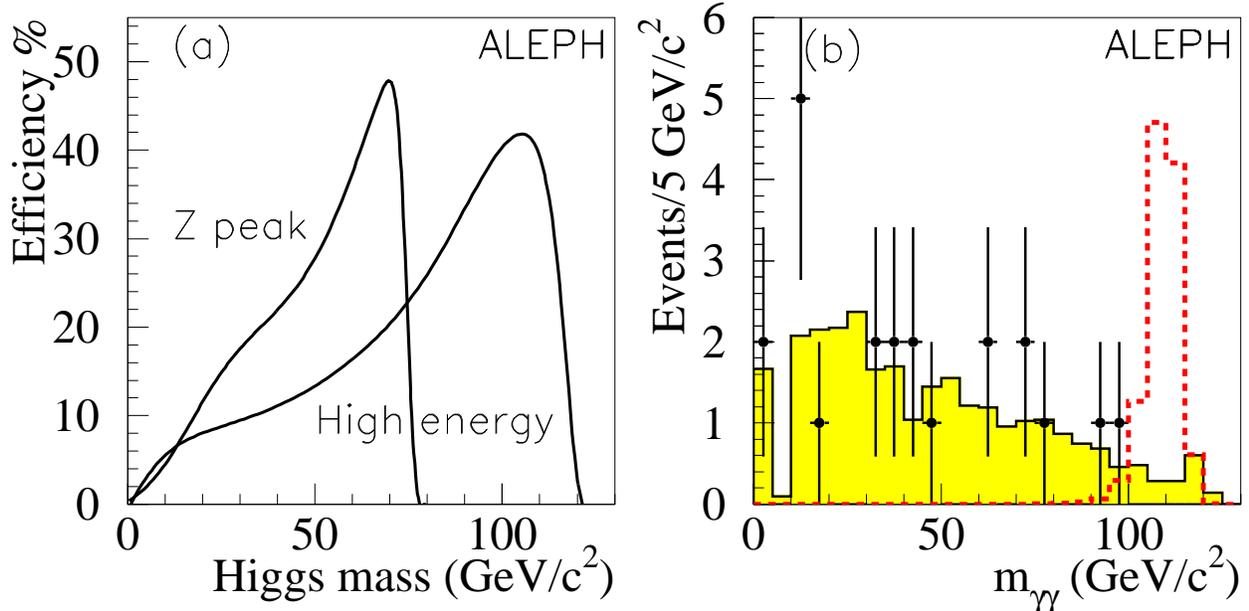}}
\caption{\protect\footnotesize\label{fig:mass_final} {(a) $\mathrm{Hf\bar{f}}$ efficiency at the Z peak
and high energy ($\mathrm{\sqrt{\it s}=202\,GeV}$).
(b) Diphoton invariant mass distribution
for data (dots with error bar), expected sources
of background (solid histogram) and a $\mathrm{110\,GeV/{\it c}^2}$ Higgs boson
signal
at $\mathrm{\sqrt{\it s}=202}\,GeV$ with arbitrary normalization (dashed histogram).}} 
\end{figure}

The signal efficiencies
are displayed in Fig.~\ref{fig:mass_final}a
as a function of the Higgs boson mass hypothesis.
Twenty-two events are selected in the data compared with an expectation of $27.9\pm 1.6$
events from all standard model background processes. The contributions of
the various channels are given in Table~\ref{tab:summary}. The
diphoton invariant mass distribution of these selected events is shown
for both data and Monte Carlo expectation in Fig.~\ref{fig:mass_final}b.
The low mass background is dominated by events where both photons
originated from the
same jet, whereas the high mass background is dominated by events where
the two photons originated from different jets. 
The sum of these two contributions has a minimum at around $\mathrm{8\,GeV/{\it c}^2}$.
No evidence of a resonance
decaying to $\gamma\gamma$ is observed and a $95\%$ confidence level upper limit on the number
of signal events at a given diphoton invariant mass is derived following the
method described in Ref.~\cite{kn:ref6}. 
For each LEP energy, the
total efficiency and the mass resolution are parametrized as a function
of the $\gamma\gamma$ invariant mass. The diphoton invariant mass resolution varies linearly
from $1\ \mathrm{GeV/{\it c}^2}$ to $3.5\ \mathrm{GeV/{\it c}^2}$ over the whole Higgs
boson mass range. The systematic uncertainties are
conservatively taken into account by
scaling down the efficiency by $5.2\%$ and increasing the
mass resolution as described in the previous section. 

\begin{figure}[!ht]
\vspace*{-1.1cm}

\centerline{\epsfxsize 14.5 truecm\epsfbox[0 90 567 567]{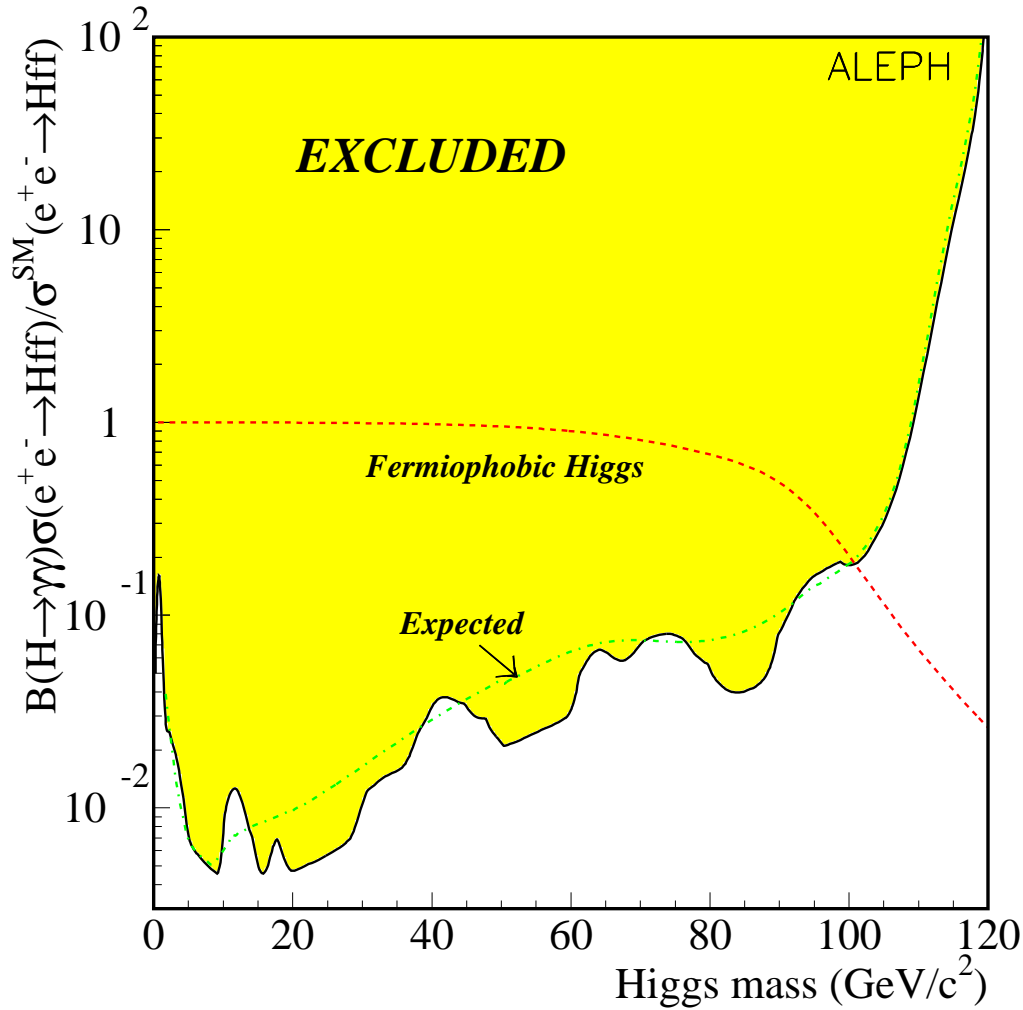}}
\vspace*{1.9cm}
\caption{\protect\footnotesize\label{fig:limit1} {  Measured (full curve) and
expected (dash-dotted
curve) $95\%$ confidence level upper limit
on $\mathrm{B(H\rightarrow\gamma\gamma)\sigma(e^+e^-\to Hf\bar{f})/\sigma^{SM}(e^+e^-\to Hf\bar{f})}$.
The dashed curve is the predicted branching ratio for a fermiophobic
Higgs boson in the limit of $\mathrm{B(H\rightarrow f\bar{f})}=0$.}}
\end{figure}
 
The present analysis does not apply for
Higgs boson masses below 1 $\mathrm{GeV/{\it c}^2}$. To extend the search to smaller masses,
the direct measurement of the Z invisible width with single photon
counting described in Ref.~\cite{kn:single} was used.
For this measurement, the single photon candidate events were selected as final states
with only one cluster in the electromagnetic calorimeter.
The $\mathrm{H\nu\bar{\nu}}$ events with $\mathrm{m_{H}< 1\,GeV/{\it c}^2}$
would therefore have been selected with a good efficiency, ranging from 13 to
$45\%$, thus leading to a sensitivity similar to that of the present analysis.

Assuming a Higgs boson production cross section with the same 
$\mathrm{\sqrt{\it s},m_H}$ dependencies as in the standard model,
the $95\%$ confidence
level upper limit on the product branching ratio
$\mathrm{B(H\to\gamma\gamma)\sigma(e^+e^-\to Hf\bar{f})/\sigma^{SM}(e^+e^-\to
Hf\bar{f})}$
is derived and
shown in Fig.~\ref{fig:limit1}.

For the case of a Higgs boson produced at the standard model rate,
the best upper limit on the branching ratio (4.7$\times 10^{-3}$ 
at $95\%$ confidence level) is obtained for Higgs boson masses below
20~$\mathrm{GeV/{\it c}^2}$.
A Higgs boson decaying exclusively to two photons is ruled out up to 
109~$\mathrm{GeV/{\it c}^2}$ at $95\%$ confidence level.

A fermiophobic Higgs boson~\cite{kn:dmc8} with no tree-level coupling to fermions is
excluded at $95\%$ confidence level for any mass up to 100.7 $\mathrm{GeV/{\it c}^{2}}$.

The present analysis extends the reach of similar analyses performed by other
LEP collaborations~\cite{kn:delphi,kn:L3_1,kn:opal}.

\section{Conclusion} 
With a data sample of 672 $\mathrm{pb}^{-1}$ recorded at centre-of-mass
energies from 88 GeV to 202~GeV, a search for two photon decays
of Higgs bosons produced in association with a fermion pair has
been performed in the mass range from 0 up to 120 $\mathrm{GeV/{\it c}^2}$.
No evidence for resonant production of photon
pairs has been found. In the framework of Higgs bosons with standard model
coupling to gauge bosons, a $95\%$ confidence level upper limit
on the Higgs boson branching ratio to two photons has been obtained
for Higgs boson masses from 0 to 109 $\mathrm{GeV/{\it c}^2}$.
In the fermiophobic model in which the Higgs boson couples 
exclusively to gauge bosons, a $95\%$
confidence level lower limit on the Higgs boson mass has been set at 100.7
$\mathrm{GeV/{\it c}^2}$.

\section{Acknowledgements}
We wish to thank our colleagues from the CERN accelerator divisions for the
successful operation of LEP. We are indebted to the engineers and technicians
in all our institutions for their contribution to the excellent performance
of ALEPH. Those of us from non-member countries thank CERN for its hospitality.

\end{document}